\documentclass[aip,amsmath,amssymb,reprint]{revtex4-1}
\usepackage{graphicx}
\usepackage{dcolumn}
\usepackage{bm}

\usepackage[utf8]{inputenc}
\usepackage[T1]{fontenc}
\usepackage{mathptmx}
\usepackage{etoolbox}

\makeatletter
\def\@email#1#2{%
  \endgroup
  \patchcmd{\titleblock@produce}
   {\frontmatter@RRAPformat}
   {\frontmatter@RRAPformat{\produce@RRAP{*#1\href{mailto:#2}{#2}}}\frontmatter@RRAPformat}
   {}{}
}%
\makeatother
\begin{document}

\preprint{AIP/123-QED}

\title[Reversible to Irreversible Transitions for Cyclically Driven Particles on Periodic Obstacle Arrays]{Reversible to Irreversible Transitions for Cyclically Driven Particles on Periodic Obstacle Arrays  
} 
\author{C. Reichhardt and C. J. O. Reichhardt}
\email{cjrx@lanl.gov}
\affiliation{
Theoretical Division and Center for Nonlinear Studies,
Los Alamos National Laboratory, Los Alamos, New Mexico 87545, USA
}%

\date{\today}

\begin{abstract}
We examine the collective dynamics of disks moving through a square array of obstacles under cyclic square wave driving. Below a critical density we find that system organizes into a reversible state in which the disks return to the same positions at the end of every drive cycle. Above this density, the dynamics are irreversible and the disks do not return to the same positions after each cycle. The critical density depends strongly on the angle $\theta$ between the driving direction and a symmetry axis of the obstacle array, with the highest critical densities appearing at commensurate angles such as $\theta=0^\circ$ and $\theta=45^\circ$ and the lowest critical densities falling at $\theta = \arctan(0.618)$, the inverse of the golden ratio, where the flow is the most frustrated. As the density increases, the number of cycles required to reach a reversible state grows as a power law with an exponent near $\nu=1.36$, similar to what is found in periodically driven colloidal and superconducting vortex systems. 
\end{abstract}
\maketitle

\section{Introduction}
The transition from reversible to irreversible dynamics
was first studied in cyclically sheared dilute colloidal systems 
where the particles undergo only contact interactions with each other
\cite{Pine05}.
For
small shear amplitudes, the particles
return to the same
positions at the end of each
shearing cycle,
while for larger shear amplitudes or
higher particle densities,
the particles do not come back to the same positions
but move irreversibly, with diffusion occurring
both parallel and perpendicular to the shearing direction.
Additional studies demonstrated that such
systems are generally always in an irreversible state initially, but can
organize to a reversible state after a number of cycles which
diverges near the critical shear amplitude or density, 
suggesting that the reversible (R) to irresistible (IR) transition  
is an example of a nonequilibrium phase transition \cite{Corte08}.
For dilute 
particles,
the transition to the reversible state is called random organization since the
system forms a disordered
configuration in which collisions between particles do not occur. 
In additional studies
of periodically driven dilute colloidal
particles, it was argued that these randomly organized
states exhibit hyperuniformity \cite{Tjhung15,Weijs15,Hexner15,Lei19}. 
Transitions to irreversible states
have also been studied in other periodically driven
systems including  granular matter \cite{Royer15,Nagasawa19}  
and amorphous solids
\cite{Regev13,Fiocco14,Keim14,Regev15,Priezjev16,Leishangthem17,Khirallah21}, 
where the particles are always in contact and
the transition occurs for a critical
shear amplitude. Such systems
can also display reversibility
spanning multiple cycles
\cite{Regev13,Lavrentovich17,Khirallah21,Keim21} as well as a variety of 
memory effects \cite{Fiocco14,Paulsen14,Keim19,Mungan19}. 

In another class of systems of collectively interacting
particles that is cyclically driven
over quenched disorder \cite{Reichhardt17},
such as vortices in type-II superconductors
\cite{Mangan08,Okuma11,Pasquini21,Maegochi21,Maegochi19},
magnetic skyrmions \cite{Litzius17,Brown18}, and
colloidal particles \cite{Stoop18},
the drive is applied uniformly to all the particles,
and in the absence
of quenched disorder,
the assembly moves back and forth uniformly in a reversible
manner.  When quenched disorder is present, however,
plastic deformations can occur that permit particles to move relative
to one another from cycle to cycle
\cite{Reichhardt17}. 
In vortex systems, studies of
R-IR transitions in 
simulation \cite{Mangan08} and experiments
\cite{Okuma11,Pasquini21,Maegochi21,Maegochi19},
show that diverging time scales appear
near critical drive amplitudes  \cite{Maegochi21}
and critical densities \cite{Maegochi19} with
critical exponents similar to those observed in
the dilute colloidal systems. 
Stoop {\it et al.} \cite{Stoop18} applied backward and forward pulse driving to
hard sphere colloids
moving over random obstacle arrays 
and obtained a variety of different dynamical phases as a function of obstacle density. 

R-IR transitions can
also occur for particles moving over a periodic array of obstacles
or a periodic substrate \cite{Reichhardt17}.
The dynamics of particles coupled to periodic substrates has been explored
for superconducting vortices \cite{Baert95, Harada96,Maegochi19,Gutierrez09}, 
magnetic skyrmions \cite{Reichhardt15a,Feilhauer20},
and colloidal systems
\cite{Korda02,MacDonald03,Bohlein12,Juniper16,Cao19,Stoop20}.  
A key aspect of systems with periodic substrates is that the
dynamics depend strongly on the direction $\theta$ of drive
relative to a substrate symmetry direction.
For example, 
in a square obstacle array, at $\theta=0^\circ$
the particles can flow easily between the
obstacles without collisions. Similarly, other drive angles  
such as $\theta=45^\circ$ and $\theta=90^\circ$
are also aligned with easy flow directions \cite{Reichhardt99,Korda02}. 
At incommensurate angles, particles cannot easily travel in a straight line without
encountering an obstacle, and the flow is more disordered. 
As the direction of a dc drive is changed relative
to the substrate,
a series of directional or symmetry locking effects
appear in which the particle
motion becomes locked to certain symmetry
directions of the substrate  even when
the drive  is not aligned precisely along  those directions 
\cite{Korda02,Reichhardt15a,Cao19,Reichhardt20,Stoop20}. 
This directional locking effect has been studied as a method for particle
separation \cite{MacDonald03,Li07,McGrath14,Wunsch16} 
and
in the context of
transitions from ordered to disordered flow
\cite{Reichhardt99,Reichhardt21,Stoop20}. 
In previous work on dc driven disks moving though square obstacle arrays,
it was shown
that the system is susceptible to jamming
for flow along certain non-symmetry angles \cite{Nguyen17,Reichhardt21}. 

Particles cyclically driven over a periodic substrate array provide
a convenient system
in which to study reversible to irreversible transitions
since the effective frustration of the array can be tuned
simply by changing the orientation of the
drive relative to the symmetry directions of the array. 
In this work we examine a monodisperse
assembly of disks interacting with a square obstacle array 
under periodic square wave driving.
When the drive is applied along
$\theta=0^\circ$ or $\theta=45^\circ$, the system 
readily organizes to a reversible pattern forming state in which
the particles return to the same positions after each drive cycle.
For driving at incommensurate angles with a fixed drive
amplitude, we find
that there is a critical disk density above
which an irreversible state forms that exhibits diffusive dynamics.
Below this density, the number of cycles required to reach a reversible state
varies with density as as a power law
with the same exponents found for periodically sheared
colloidal particles \cite{Corte08} and driven
superconducting vortex systems  \cite{Okuma11,Pasquini21,Maegochi21,Maegochi19}.
In general we find that the disks form a disordered or fluid like state 
under irreversible flow,
while a pattern forming or ordered configuration appears when the
system reaches a reversible state. 
The critical density for the R-IR transition is nonmonotonic as a function
of $\theta$,
reaching maximum values for
commensurate driving angles and showing a global minimum
near the
arctangent of the inverse of the golden ratio.
Our results could be tested in a variety of systems
such as colloidal particles, superconducting vortices, or magnetic skyrmions
under periodic driving coupled to a periodic array of obstacles
or a periodic substrate.

\section{Simulation}
We consider a two dimensional system of size $L \times L$
containing a square array of $N_{\rm obs}=81$ circular obstacles of lattice spacing $a$ and 
radius $r_{\rm obs}$. We impose periodic
boundary conditions in the $x$ and $y$-directions and 
place $N_{d}$ monodisperse repulsive disks in the sample.
The dynamics of disk $i$
is governed by the following overdamped equation of motion: 
\begin{equation} 
\alpha_d {\bf v}_{i}  =
{\bf F}^{dd}_{i} +  {\bf F}^{obs}_{i} + {\bf F}^{D}  \ .
\end{equation}
The velocity of the disk at position ${\bf r}_i$ is 
${\bf v}_{i} = d {\bf r}_{i}/dt$ and
we set the damping constant $\alpha_d$  to unity.  
The first term on the right is the disk-disk interaction force 
${\bf F}^{dd}_{i}$ 
represented by a short-range harmonic repulsive potential with radius $r_{d}$.
In this work we fix $r_d=0.55$.
The disk-obstacle force ${\bf F}^{\rm obs}_i$ is also modeled as a repulsive 
harmonic interaction.
In our work we choose harmonic spring constants
that are large enough to prevent the overlap
between disks from becoming larger than one percent
for the densities and driving forces we consider.
The density $\phi$ is
defined to be the area covered by the obstacles and mobile disks, 
$\phi = N_{\rm obs}\pi r^2_{obs}/L^2 + N_{d}\pi r^2_{d}/L^2$.
The square wave driving force remains constant for
a fixed period of time in the forward direction prior to reversing, and has the form
${\bf F}_{D} = A\cos(\theta){\bf \hat x} + A\sin(\theta){\bf \hat y}$,
where $A$ is the drive amplitude and $\theta$ is the
direction of the drive relative to the $x$-axis of the periodic array.  
This model was previously
employed to study locking and clogging effects for
dc driven disks \cite{Reichhardt20,Reichhardt21}.  
We fix the duration of the drive to
$T/2 = 2\times 10^5$ simulation time steps spent on each half cycle,
and we vary the density $\phi$ and the drive amplitude $A$. 
We can characterize the system by measuring the change in the
position of the disks from one cycle to the next,
$R(n) = \sum^{N_{d}}_{i}[{\bf r}_{i}(t_{0} + nT) - {\bf r}_{i}(t_{0} + (n-1)T)]$,
where $t_0$ is an initial reference time.
If the motion
is reversible, $R_{n}=0$.
We also measure the total net displacement $d(n)$
as a function of cycle $n$,
$d(n) = \sum^{N_{d}}_{i}[{\bf r}_{i}(t_{0} + nT) - {\bf r}_{i}(t_{0})]$.
In an irreversible state, $d(n)$  grows continuously,
while for a reversible state,
$d(n)$ saturates to a finite value.  

\section{Results}

\begin{figure}
\includegraphics[width=\columnwidth]{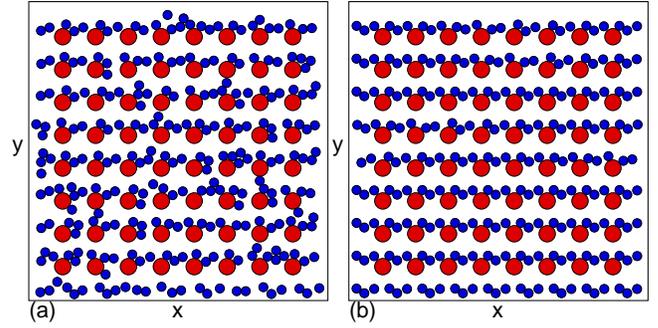}
\caption{
Snapshots of the obstacle locations (red) and mobile disks (blue) in a portion
of a sample
with $r_{\rm obs} = 1.0$.
The amplitude of the square wave drive
is
$A=0.031623$ and the driving angle is $\theta=18.435^\circ$.
(a)
An irreversible state at $N_d=269$ and $\phi=0.3962$.
(b) A reversible state at $N_d=239$ and $\phi  = 0.3716$ .
}
\label{fig:1}
\end{figure}

In Fig.~\ref{fig:1}(a) we show a snapshot of the obstacle locations
and mobile disks 
for a system with $r_{\rm obs} = 1.0$,
$A=0.031623$ and a driving angle of $\theta=18.435^\circ$ for
$N_{d} = 269$,
giving an overall system density of $\phi = 0.3962$.
Even after 1500 ac drive cycles, the system remains in an irreversible or
fluctuating state. When the number of mobile disks is reduced to
$N_d=239$, giving $\phi=0.3716$,
the system organizes into a reversible state
with an ordered structure, as illustrated in 
Fig.~\ref{fig:1}(b).

\begin{figure}
\includegraphics[width=\columnwidth]{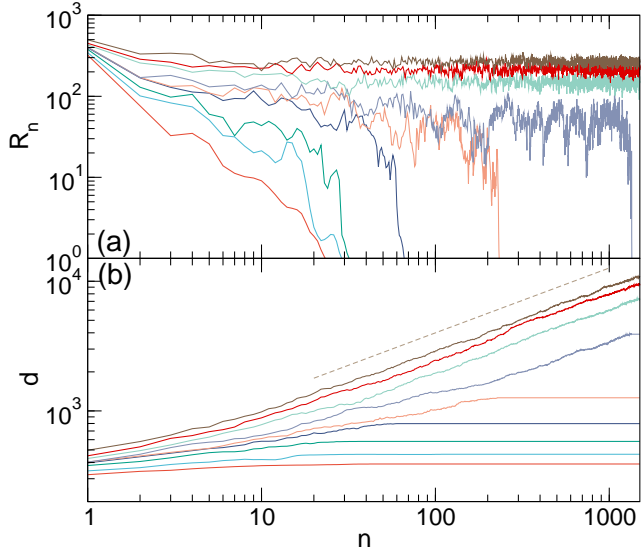}
\caption{
(a) $R_n$ versus cycle number $n$ for the system
in Fig.~\ref{fig:1} with $r_{\rm obs}=1.0$ 
for increasing total density
$\phi = 0.335$, 0.3496, 0.3569, 0.36427, 0.36867, 0.3716, 0.3789, 0.3862,
and  $0.3962$, from top to bottom.
Here $A=0.031623$ and $\theta=18.435^\circ$.
For  $\phi < 0.3789$, $R_{n}$ goes to zero,
indicating the system has reached a reversible state.
(b) The corresponding $d$ versus $n$. For $\phi > 0.3962$,
$d$ grows continuously, while at densities smaller than this, $d$ saturates
to a finite value.
The dashed line is a fit to $d \propto n^{1/2}$.
}
\label{fig:2}
\end{figure}

In Fig.~\ref{fig:2}(a) we plot $R_n$ versus cycle number $n$
for the system in Fig.~\ref{fig:1} for increasing total densities of
$\phi = 0.335$, 0.3496, 0.3569, 0.36427, 0.36867, 0.3716, 0.3789, 0.3862,
and  $0.3962$.
For  $\phi < 0.3789$, $R_{n}$ goes to zero, indicating that after
an initial transient of some length,
the disks return to the same positions after every driving cycle
and the system behaves reversibly. As
$\phi$ increases, the number of cycles $\tau$
required to reach a reversible state also increases.
For example, at $\phi = 0.3716$ it takes
$\tau=1360$ cycles to reach the reversible state
illustrated in Fig.~\ref{fig:1}(b), while at lower densities
such as $\phi= 0.335$, $\tau=12$.
Figure~\ref{fig:2}(b) shows the corresponding $d$
versus $n$. For $\phi > 0.3962$, 
$d$ continues to grow as a function of time,
while below this density it saturates to a finite value.
The dashed line 
indicates a fit to $d \propto n^{1/2}$.
Since $n$ also corresponds to an elapsed time,
this implies that the displacements
are growing as $t^{1/2}$ and thus have
Brownian characteristics,
similar to the behavior of the displacement
found in the irreversible states of
sheared colloidal systems \cite{Pine05}. 

\begin{figure}
\includegraphics[width=\columnwidth]{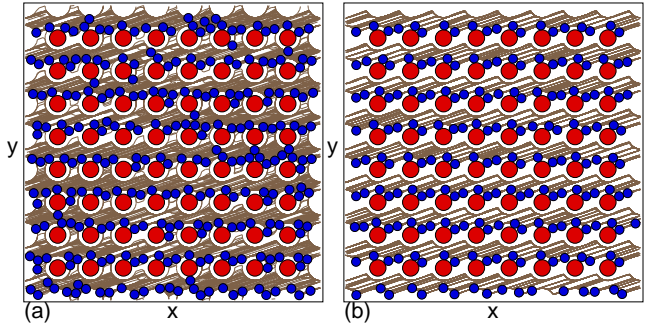}
\caption{
Obstacle locations (red), mobile disks (blue), and mobile disk trajectories
(brown)  
in a portion of the sample
for the system in Fig.~\ref{fig:2} with $r_{\rm obs}=1.0$, 
$A=0.031623$, and $\theta=18.435^\circ$.
(a) An irreversible state at $\phi = 0.3962$,
where the trajectories gradually fill all of space and 
the disks undergo long time diffusion.
(b) A reversible state at $\phi = 0.3496$, where the disks
repeatedly follow the same path
and the motion is confined between the rows of obstacles.
}
\label{fig:3}
\end{figure}

In Fig.~\ref{fig:3} we plot the trajectories of the disks from the system in 
Fig.~\ref{fig:2} to   
illustrate more clearly the difference between the irreversible 
and reversible dynamics. 
Figure~\ref{fig:3}(a) shows
that the trajectories in the irreversible state
at $\phi = 0.3962$
fill space, and the disks are translating in both
the $x$ and $y$ directions. 
In Fig.~\ref{fig:3}(b),
the reversible state at $\phi = 0.3496$
contains much more ordered trajectories and
the motion is always confined between rows of obstacles
with no hopping from row to row.
If the trajectory plot in the reversible state is extended over a larger number of cycles,
exactly the same same trajectory pattern appears.

\begin{figure}
\includegraphics[width=\columnwidth]{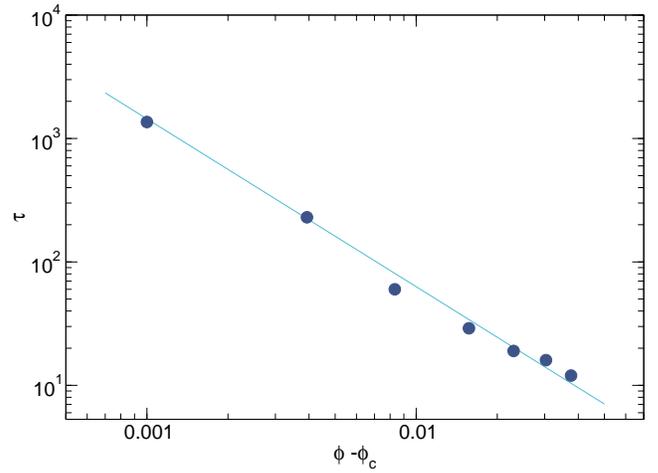}
\caption{
The number of cycles $\tau$ required to reach the reversible state
versus $\phi-\phi_c$, where we have assumed a critical density of  
$\phi_{c} = 0.3726$, for the system from Fig.~\ref{fig:2} with
$r_{\rm obs}=1.0$, 
$A=0.031623$, and $\theta=18.435^\circ$.
The line is a fit to
$\tau =  (\phi -\phi_{c})^{-\nu}$ with $\nu = 1.36$.
}
\label{fig:4}
\end{figure}

In Fig.~\ref{fig:4} we plot the the number of cycles
$\tau$ required to reach the reversible state as a function of
$\phi-\phi_c$,
where we assume a
critical density of $\phi_{c} = 0.3726$. The line indicates a fit to
$\tau =  (\phi -\phi_{c})^{-\nu}$ with $\nu = 1.36$.
Previous 
work on R-IR transitions in 
two-dimensional (2D)
sheared colloidal systems showed
a similar divergence in the time to reach the reversible state
with 
$\nu =1.33$ \cite{Corte08}, while studies of
superconducting vortices driven over random disorder
gave exponents of
$\nu=1.38$ for critical drive amplitudes
and $\nu=1.32$ for critical densities\cite{Maegochi21}.
These exponents are 
close to those expected for
2D directed percolation, where $\nu=1.295$ \cite{Hinrichsen00}.  

The images in Fig.~\ref{fig:3} clarify
why the R-IR transition
is connected to percolation. In a reversible state, the
trajectories do not simultaneously percolate in both the $x$
and $y$ directions,
while in an irreversible state, the trajectories are mixing. 
The percolation transition
could be considered to occur at the point
where the trajectories just begin to overlap in both
the $x$ and $y$ directions.

\begin{figure}
\includegraphics[width=\columnwidth]{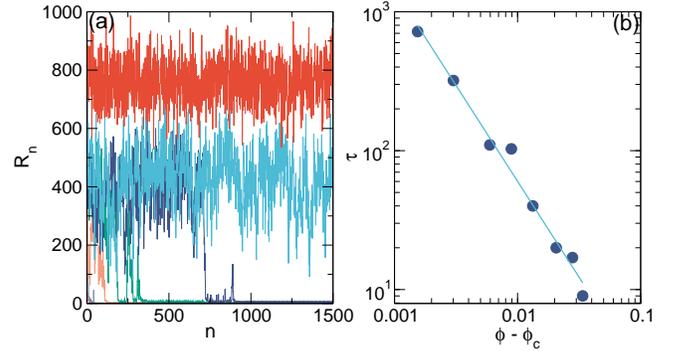}
\caption{
$R_{n}$ versus $n$ for the system in
Fig.~\ref{fig:2} with $r_{\rm obs}=1.0$
and $A=0.031623$
but for driving along $\theta = 45^\circ$ or a commensurate
angle.
The total density is $\phi = 0.5124$, 0.5183, 0.53295, 0.5402, 0.54319,
0.5446, 0.54612, and  $0.56225$, from bottom to top.
(b) The corresponding $\tau$ versus $\phi-\phi_c$ where
we assume a critical density of $\phi_c=0.54612$. The line is a fit to
$\tau =  (\phi -\phi_{c})^{-\nu}$ with $\nu = 1.38$. 
}
\label{fig:5}
\end{figure}

\begin{figure}
\includegraphics[width=\columnwidth]{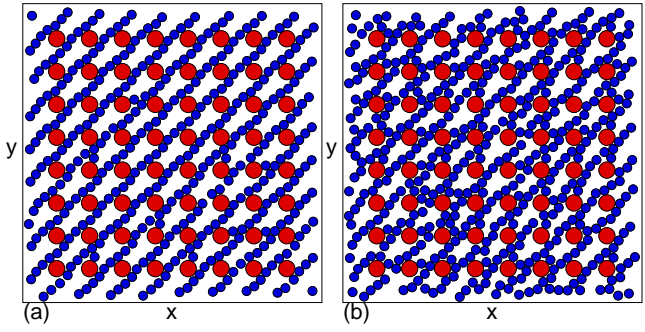}
\caption{
Snapshots of the obstacle locations (red) and mobile disks (blue) in a portion
of the sample
from Fig.~\ref{fig:4} with $r_{\rm obs}=1.0$, 
$A=0.031623$, and $\theta=45^\circ$.
(a) A reversible state at $\phi = 0.5036$ where chains
form that are aligned with the driving direction of $\theta=45^\circ$.
(b) An irreversible state at $\phi = 0.6$ where the configuration is disordered.
}
\label{fig:6}
\end{figure}

We next consider the R-IR transition
for driving at commensurate angles.
In  Fig.~\ref{fig:5} we plot $R_{n}$ vs $n$ for the same system in
Fig.~\ref{fig:2} but at a driving angle of
$\theta = 45^\circ$ for
total densities of $\phi = 0.5124$ to $0.56225$.
When $\phi<0.5446$,
the system organizes to a reversible state in
a time $\tau$ that grows with increasing $\phi$.
In Fig.~\ref{fig:4}(b) we plot
$\tau$ versus $\phi-\phi_c$ for a critical density of
$\phi_c=0.54612$, as well as a line indicating a fit to
$\tau =  (\phi -\phi_{c})^{-\nu}$ with $\nu = 1.38$.
This result indicates that the R-IR transition shown in Fig.~\ref{fig:4}
persists for driving
along $\theta=45^\circ$; however, the 
critical density $\phi_c$ is higher.
In Fig.~\ref{fig:6} we illustrate the disk configurations
above and below the critical R-IR transition density.
Figure~\ref{fig:6}(a) shows a reversible state at $\phi = 0.5036$
where the disks form ordered one-dimensional (1D) chains
aligned with the drive
along $\theta=45^\circ$.
In Fig.~\ref{fig:6}(b), the same system at
$\phi = 0.577$
is in a irreversible state where the disk positions are disordered. 

\begin{figure}
\includegraphics[width=\columnwidth]{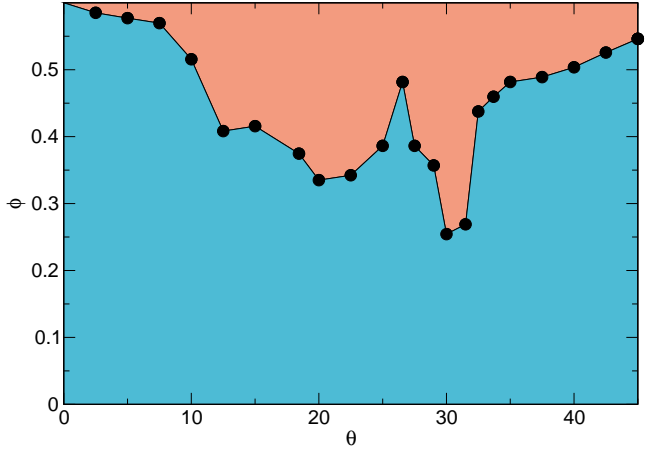}
\caption{Phase diagram as a function of
total density $\phi$ versus driving angle $\theta$
showing regions where the system organizes to a reversible state (blue)
or an irreversible state (orange)
for the system in Fig.~\ref{fig:2} with $r_{\rm obs}=1.0$
and $A  = 0.031623$.
Circles indicate the transition density $\phi_c$.
The reversible region reaches maximum extents at
$\theta=0^\circ$, $\theta=45^\circ$, and
$\theta=26.5^\circ=\arctan(1/2)$.
The minimum width of the reversible region falls
near $\theta = 31.0^\circ$ or close to 
$ \theta = \arctan(0.618)$ where $0.618$ is the inverse of the golden ratio. 
}
\label{fig:7}
\end{figure}

By conducting a series of simulations for fixed
driving amplitude $A  = 0.031623$
and varied $\theta$, we explore the dependence of $\phi_c$
on the driving angle $\theta$.
We plot the reversible and irreversible regions
as a function of $\phi$ versus $\theta$ in
Fig.~\ref{fig:7}.
Note that for symmetry reasons, the pattern shown in Fig.~\ref{fig:7} repeates in an
inverted fashion over the range $\theta=45^\circ$ to $\theta=90^\circ$.
At $\theta = 0^\circ$, the system remains in a reversible state
up to the largest values of $\phi$ we consider, $\phi=0.61$.
For larger densities, jamming effects become important and we would
need to switch to a different disk initialization algorithm.
It may be possible that additional R-IR transitions occur at higher
disk densities when jammed states begin to appear; however,
this is beyond the scope of the present work.
For $0 < \theta < 7.5^\circ$, we find that
the R-IR transition occurs
near a critical density $\phi_c = 0.575$.
There is a peak in $\phi_c$ 
near $\theta=26.565^\circ$, which 
corresponds to a commensurate angle of
$\theta = \arctan(1/2)$.
When the driving angle is close 
to lattice symmetry directions such as $0/1, 1/2$, or $1/1$,
which correspond
to $\theta=0^\circ$, $26.5^\circ$, and $45^\circ$, respectively,
$\phi_c$ reaches its highest values.
Under these commensurate angles,
the disks can move easily along straight lines
while avoiding collisions with the obstacles.
There is no noticeable peak
in $\phi_c$ at $\theta=\arctan(1/3)$ or
$\theta=\arctan(2/3)$, and the disk dynamics
for these driving angles are similar to what is found at
incommensurate driving angles.  
A minimum in $\phi_c$
occurs near
$\theta = 31^\circ$ or close 
to $ \theta = \arctan(0.618)$, where $0.618$
is the inverse of the golden ratio 
from the Fibonacci sequence.
The incommensuration is maximized 
at the inverse golden ratio
where the driven disk collides with the
largest possible number of obstacles while moving through the system.
The variations in the extent of the reversible regions
should also depend on the radius $r_d$ of the mobile disk.
If a smaller disk were
used, other possible commensuration effects could
appear depending on
how many rows of mobile particles can fit along $45^\circ$ or other
commensurate angles.

\begin{figure}
\includegraphics[width=\columnwidth]{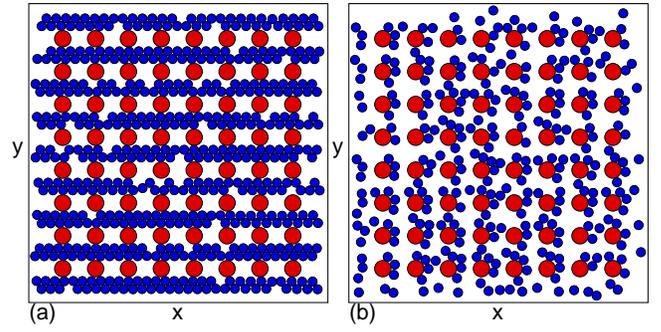}
\caption{
Snapshots of the obstacle locations (red) and mobile disks (blue) in a portion
of the sample from Fig.~\ref{fig:7} with $r_{\rm obs}=1.0$ 
and $A=0.031623$.
(a) A reversible state
at $\theta = 0^\circ$ and
$\phi = 0.577$,
where the system forms a pattern of chains aligned along the $x$ direction.
(b) A reversible state at $\theta = \arctan(1/2) = 26.5^\circ$ 
for $\phi = 0.4302$, where the repeating pattern is disordered.
}
\label{fig:8}
\end{figure}

In Fig.~\ref{fig:8}(a) we illustrate the disk positions in
a reversible state at $\theta =  0^\circ$ and 
$\phi =  0.577$, where the disks
form two nearly filled rows
moving in the $x$ direction.
We note that not all of the reversible states
are associated with ordered disk arrangements.
For example, at $\theta = \arctan(1/2)$,
where a peak in $\phi_c$ appears in Fig.~\ref{fig:7},
the
system forms the disordered but repeatable pattern
shown in Fig.~\ref{fig:8}(b) for $\phi = 0.4302$ in the reversible state.  

\begin{figure}
\includegraphics[width=\columnwidth]{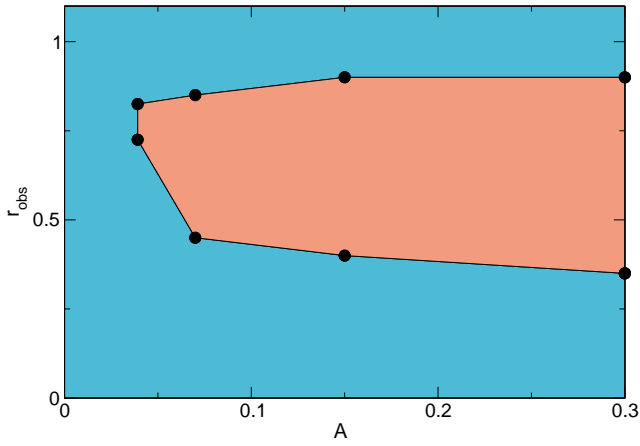}
\caption{
Phase diagram as a function of obstacle radius $r_{\rm obs}$ versus
driving amplitude $A$ showing the reversible (blue) and irreversible
(orange) regimes
for a system with $N_d=279$
at $\theta=7.5^\circ$.
Here, when $r_{\rm obs} = 1.0$, $\phi = 0.041$.
There is a reentrant R-IR transition 
as a function of $r_{\rm obs}$ for all but the lowest values of $A$.
}
\label{fig:9}
\end{figure}

Up to this point we have concentrated on samples with
$r_{\rm obs} = 1.0$,
but there can be a reentrant R-IR transition as
$r_{\rm obs}$ is varied.
In a system with $\theta = 7.5^\circ$, Fig.~\ref{fig:7} indicates that an irreversible
state appears only for large values of $\phi$. In this case, a transition occurs from
1D reversible motion of disks along the $x$ direction
to 2D irreversible motion.
If we reduce $r_{\rm obs}$, the transition to irreversible motion shifts
to lower $\phi$ because
the disks can more readily move in two dimensions instead of
remaining locked in a 1D channel.
If, however, $r_{\rm obs}$ is reduced even further,
collisions with the obstacles become less frequent
and the system can once again organize
into a reversible state.
This is illustrated in Fig.~\ref{fig:9} where
we plot the locations of the reversible
and irreversible regimes as a function of
$r_{\rm obs}$ versus $A$ for a system with 
$N_d=279$
at $\theta=7.5^\circ$.
Here, when $r_{\rm obs} = 2.0$, $\phi = 0.401$.
For large $r_{\rm obs} > 0.9$, the
system is always in a reversible state regardless of the value of $A$,
and the disks form 1D chains.
Furthermore, for $A < 0.04$
the system is 
always in a reversible state since the disks do not move far enough during
a single drive cycle to collide with the obstacles.
When $A > 0.04$ and  $0.4 < r_{\rm obs} < 0.9$,
irreversible behavior appears,
while for $r_{\rm obs} < 0.4$,
there is a reversible state in which
the disks are moving.
The result is the apperance of a reentrant R-IR transition as a function of $r_{\rm obs}$
for all but the smallest values of $A$.
Similar reentrant transitions
should occur near commensurate driving
angles such as $\theta =45^\circ$. In contrast, for
incommensurate
angles the system will
remain in an irreversible
state down to much smaller $r_{\rm obs}$ 
since even relatively small
moving disks continue to collide with the obstacles due to the driving direction.
The existence of reentrance 
will also depend on the mobile disk density since
for low
mobile disk densities the system will generally
be able to organize into a reversible state.

\begin{figure}
\includegraphics[width=\columnwidth]{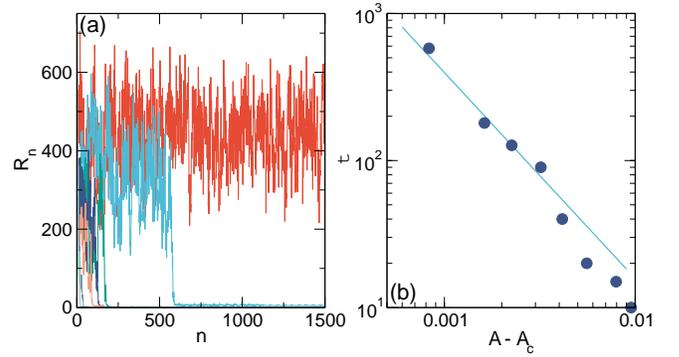}
\caption{
$R_{n}$ versus $n$ for a
system with $r_{\rm obs}=1.0$
and $\phi=0.549$
for driving along
the commensurate angle $\theta = 45^\circ$.
When $A = 0.03162$, this system is in an irreversible state.
From top to bottom, $A = 0.02846$, 0.02767, 0.0268, 0.025247, 0.0253, 0.02435, 
and $0.022136$.  
(b) The corresponding $\tau$ versus $A-A_c$ where we assume a critical
drive of $A_c=0.0285$.
The line is a fit to
$\tau = |A - A_{c}|^{-\nu}$
with $\nu = 1.4$.
}
\label{fig:10}
\end{figure}

We can also observe a R-IR transition at fixed $\phi$ under
increasing $A$,
as illustrated in Fig.~\ref{fig:10} for a system with
$\phi  = 0.549$ and $\theta = 45^\circ$.
When $A = 0.03162$, this system is in an irreversible state.
In Fig.~\ref{fig:10}(a) we plot
$R_n$ versus $n$ for $A = 0.02846$,
0.02767, 0.0268, 0.025247, 0.0253, 0.02435,
and $0.022136$.
The system organizes to a reversible
state when $A < 0.02846$,
and the number of cycles $\tau$ needed to reach the reversible state
decreases with decreasing $A$. 
In Fig.~\ref{fig:10}(b) we plot
$\tau$ versus $A-A_c$ where $A_c=0.0285$. The line is a fit to
$\tau = |A - A_{c}|^{-\nu}$
with $\nu = 1.4$,
an exponent slightly larger than what we observe when varying the
total density.
The exponents are not accurate enough to determine whether the two transitions
are in different universality classes.
If we focus only on the points closest to $A_{c}$, a lower value of $\nu$ can
be fit, suggesting that both transitions in fact fall in the
same universality class.  

\section{Summary}
We have numerically examined the reversible to irreversible transition for
periodically driven disks
moving through a two-dimensional square periodic obstacle array.
For fixed ac drive amplitude, we find that there is
a critical density at which the system is able to organize into a reversible
state instead of remaining in an irreversible state.
The number of cycles required to reach the reversible state
diverges as a power law with
an exponent $\nu \approx 1.36$.
This is close to the value of $\nu$ observed for
periodically sheared colloidal particles
and periodically driven superconducting vortices, suggesting that
the reversible-irreversible transitions of all of these systems fall into
the same universality class.
The critical density at which the transition occurs
is non-monotonic as a function of the angle between the applied drive and
a symmetry direction of the obstacle array.
The highest critical densities
appear for commensurate driving angles such as
$\theta=0^\circ$ and $\theta=45^\circ$.
We find the same power law exponents 
for both incommensurate and commensurate angles.
We obtain the lowest critical density for
$\theta = \arctan(0.618)$,
which is the inverse of the golden ratio.
This frustrated driving direction produces the highest frequency of collisions
between disks and obstacles.

\begin{acknowledgments}
We gratefully acknowledge the support of the U.S. Department of
Energy through the LANL/LDRD program for this work.
This work was carried out under the auspices of the 
NNSA of the 
U.S. DoE
at 
LANL
under Contract No.
DE-AC52-06NA25396 and through the LANL/LDRD program.
\end{acknowledgments}

\section*{Data Availability Statement}
Data available on request from the authors.

\nocite{*}
\bibliography{mybib}

\end{document}